\documentclass[pra,twocolumn,superscriptaddress,showpacs,preprintnumbers,amsmath,amssymb]{revtex4}
\usepackage{bm}
\usepackage{bbm}
\usepackage{amssymb}
\usepackage{amsfonts}
\usepackage{epsfig,graphicx}
\usepackage{amstext}
\usepackage{amsmath}
\usepackage{graphicx}
\usepackage{times}
\usepackage{txfonts}
\usepackage{dcolumn}

\begin{document}

\title{Hierarchy of the nonlocal advantage of quantum coherence and Bell nonlocality}

\author{Ming-Liang Hu}
\email{mingliang0301@163.com}
\affiliation{School of Science, Xi'an University of Posts and Telecommunications, Xi'an 710121, China}
\affiliation{Institute of Physics, Chinese Academy of Sciences, Beijing 100190, China}
\author{Xiao-Min Wang}
\affiliation{School of Science, Xi'an University of Posts and Telecommunications, Xi'an 710121, China}
\author{Heng Fan}
\email{hfan@iphy.ac.cn}
\affiliation{Institute of Physics, Chinese Academy of Sciences, Beijing 100190, China}
\affiliation{CAS Center for Excellence in Topological Quantum Computation, University of Chinese Academy of Sciences, Beijing 100190, China}
\affiliation{Songshan Lake Material Laboratory, Dongguan 523000, China}

\begin{abstract}
Quantum coherence and nonlocality capture nature of quantumness from
different aspects. For the two-qubit states with diagonal
correlation matrix, we prove strictly a hierarchy between the
nonlocal advantage of quantum coherence (NAQC) and Bell nonlocality
by showing geometrically that the NAQC created on one qubit by local
measurement on another qubit captures quantum correlation which is
stronger than Bell nonlocality. For general states, our numerical
results present strong evidence that this hierarchy may still hold.
So the NAQC states form a subset of the states that can exhibit Bell
nonlocality. We further propose a measure of NAQC that can be used
for a quantitative study of it in bipartite states.
\end{abstract}


\maketitle

\section{Introduction} \label{sec:1}
Quantum correlations in states of composite systems can be
characterized from different perspectives. From the applicative
point of view, they are also invaluable physical resources which are
recognized to be responsible for the power of those classically
impossible tasks involving quantum communication and quantum
computation \cite{Nielsen}. Stimulated by this realization, there
are a number of quantum correlation measures being put forward up to
date \cite{nonlocal,entangle,steer,discord}. Some of the extensively
studied measures include Bell nonlocality (BN) \cite{nonlocal},
quantum entanglement \cite{entangle}, Einstein-Podolski-Rosen
steering \cite{steer}, and quantum discord \cite{discord}. For
two-qubit states, a hierarchy of these quantum correlations has also
been identified \cite{new01,new02,hierarchy,Angelo1,Angelo2,
Angelo3}. This hierarchy reveals different yet interlinked subtle
nature of correlations, and broadens our understanding about the
physical essence of quantumness in a state.

Quantum coherence is another basic notion in quantum theory, and
recent years have witnessed an increasing interest on pursuing its
quantification \cite{Plenio,Hu}. In particular, based on a seminal
framework formulated by Baumgratz \textit{et al.} \cite{coher},
there are various coherence measures being proposed \cite{meas6,
asym1,co-ski2,meas4,dist2,measjpa,new1}. This stimulates one's
enthusiasm to understand them from different aspects, as for
instance the distillation of coherence \cite{dist2,distill}, the
role of coherence played in quantum state merging \cite{qsm}, and
the characteristics of coherence under local quantum operations
\cite{create1,create2,mc1,mc2} and noisy quantum channels
\cite{fro1,fro2}. Moreover, some fundamental aspects of coherence
such as its role in revealing the wave nature of a system
\cite{path1,path2}, its tradeoffs under the mutually unbiased bases
\cite{comple1} or incompatible bases \cite{comple2}, have also been
extensively studied.

Conceptually, coherence is thought to be more fundamental than
various forms of quantum correlations, hence it is natural to pursue
their interrelations for bipartite and multipartite systems. In
fact, it has already been shown that coherence itself can be
quantified by the entanglement created between the considered system
and an incoherent ancilla \cite{meas1}. There are also several works
which linked coherence to quantum discord \cite{Yao,Ma,Hufan} and
measurement-induced disturbance \cite{Huxy}.

In a recent work, Mondal \textit{et al.} \cite{naqc} explored the
interrelation of quantum coherence and quantum correlations from an
operational perspective. By performing local measurements on qubit
$A$ of a two-qubit state $AB$, they showed that the average
coherence of the conditional states of $B$ summing over the mutually
unbiased bases can exceed a threshold that cannot be exceeded by any
single-qubit state. They termed this as the nonlocal advantage of
quantum coherence (NAQC), and proved that any two-qubit state that
can achieve a NAQC (we will call it the NAQC state for short) is
quantum entangled. As there are many other quantum correlation
measures, it is significant to purse their connections with NAQC. We
explore such a problem in this paper. For two-qubit states with
diagonal correlation matrix, we showed strictly that quantum
correlation responsible for NAQC is stronger than that responsible
for BN, while for general states this result is conjectured based on
numerical analysis. We hope this finding may shed some light on our
current quest for a deep understanding of the interrelation between
quantum coherence and quantum correlations in composite systems.

\section{Technical preliminaries} \label{sec:2}
We start by recalling two well-established coherence measures known
as the $l_1$ norm of coherence and relative entropy of coherence
\cite{coher}. For a state described by density operator $\rho$ in
the reference basis $\{|i\rangle\}$, they are given, respectively,
by
\begin{equation}\label{eq2-1}
 C_{l_1}(\rho)=\sum_{i\neq j}|\langle i| \rho |j\rangle|, \;
 C_{re}(\rho)=S(\rho_{\mathrm{diag}})-S(\rho),
\end{equation}
where $S(\cdot)$ denotes the von Neumann entropy, and
$\rho_{\mathrm{diag}}$ is an operator comprised of the diagonal part
of $\rho$.

Using the above measures, Mondal \textit{et al.} presented a
``steering game'' in Ref. \cite{naqc}: Two players, Alice and Bob,
share a two-qubit state $\rho$. They begin this game by agreeing on
three observables $\{\sigma_1,\sigma_2,\sigma_3\}$, with
$\sigma_{1,2,3}$ being the usual Pauli operators. Alice then
measures qubit \textit{A} and informs Bob of her choice $\sigma_i$
and outcome $a\in\{0,1\}$. Finally, Bob measures coherence of qubit
\textit{B} in the eigenbasis of either $\sigma_j$ or $\sigma_k$
($j,k \neq i$) randomly. By denoting the ensemble of his conditional
states as $\{p(a|\sigma_i), \rho_{B|\sigma_i^a}\}$, the average
coherence is given by
\begin{equation}\label{eq2-2}
 \bar{C}_\alpha^{\sigma_j}(\{p(a|\sigma_i),\rho_{B|\sigma_i^a}\})
  = \sum_a p(a|\sigma_i) C_\alpha^{\sigma_j}(\rho_{B|\sigma_i^a}),
\end{equation}
where $p(a|\sigma_i)= \mathrm{tr}(\Pi_i^a \rho)$,
$\rho_{B|\sigma_i^a}= \mathrm{tr}_A(\Pi_i^a \rho)/p(a|\sigma_i)$,
$\Pi_i^a= [I_2+ (-1)^a \sigma_i]/2$, $I_2$ is the identity operator,
and $C_\alpha^{\sigma_j}$ ($\alpha=l_1$ or $re$) is the coherence
defined in the eigenbasis of $\sigma_j$.

By further averaging over the three possible measurements of Alice
and the corresponding possible reference eigenbases chosen by Bob,
Mondal \textit{et al.} \cite{naqc} derived the criterion for
achieving NAQC, which is given by
\begin{equation}\label{eq2-3}
 C_\alpha^{na}(\rho)= \frac{1}{2}\sum_{i,j,a \atop i \neq j} p(a|\sigma_i)
                      C^{\sigma_j}_\alpha(\rho_{B|\sigma_i^a})>C_\alpha^m,
\end{equation}
where $C_{l_1}^m=\sqrt{6}$, $C_{re}^m = 3H(1/2+\sqrt{3}/6) \simeq
2.2320$, and $H(\cdot)$ stands for the binary Shannon entropy
function.

In fact, the above critical values are also direct results of the
complementarity relations of coherence under mutually unbiased bases
\cite{comple1}. To be explicit, by Eq. (4) of Ref. \cite{comple1}
and the mean inequality (the arithmetic mean of a list of
nonnegative real numbers is not larger than the quadratic mean of
the same list) one can obtain the critical value $C_{l_1}^m$, while
from Eq. (24) of Ref. \cite{comple1} one can obtain the critical
value $C_{re}^m$.

To detect nonlocality in $\rho$, one can use the Bell-CHSH
inequality $|\langle B_{\mathrm{CHSH}}\rangle_\rho| \leqslant 2$,
where $B_{\mathrm{CHSH}}$ is the Bell operator \cite{chsh}.
Violation of this inequality implies that $\rho$ is Bell nonlocal.
The maximum of $|B_{\mathrm{CHSH}} \rangle_\rho|$ over all mutually
orthogonal pairs of unit vectors in $\mathbb{R}^3$ is given by
\cite{chsh2}
\begin{equation}\label{eq2-4}
 B_{\max}(\rho) =2\sqrt{M(\rho)},
\end{equation}
where $M(\rho) = u_1+u_2$, with $u_i$ ($i=1,2,3$) being the
eigenvalues of $T^\dag T$ arranged in nonincreasing order, and $T$
stands for the matrix formed by elements $t_{ij} = \mathrm{tr}(\rho
\sigma_i \otimes \sigma_j)$. Clearly, $M(\rho)>1$ is also a
manifestation of BN in $\rho$.

\begin{figure}
\centering
\resizebox{0.41 \textwidth}{!}{%
\includegraphics{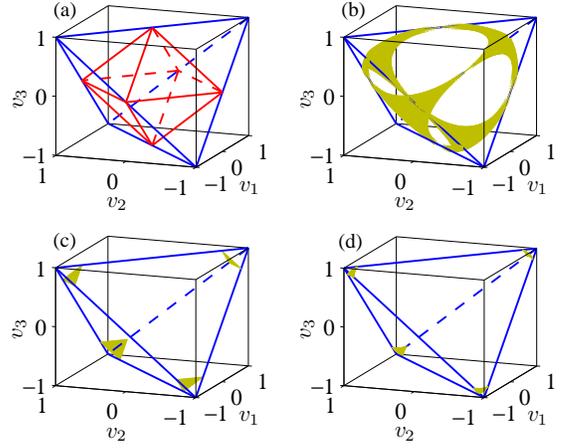}}
\caption{The tetrahedron $\mathcal{T}$ and octahedron $\mathcal {O}$
associated with $\rho_{\mathrm{Bell}}$ (a), and the level surfaces
of $M(\tilde{\rho})=1$ (b), $C_{l_1}^{na}(\rho_{\mathrm{Bell}})=
\sqrt{6}$ (c), and $C_{re}^{na}(\rho_{\mathrm{Bell}})= C_{re}^m$ (d).
The regions of Bell nonlocal states $\tilde{\rho}$ and NAQC states
$\rho_{\mathrm{Bell}}$ are those outside the level surfaces.}
\label{fig:1}
\end{figure}

It has been shown that any $\rho$ that can achieve a NAQC is
entangled, while the opposite case is not always true \cite{naqc}.
This gives rise to a hierarchy of them. To further establish the
hierarchy between NAQC and BN, and based on the consideration that
the BN is local unitary invariant, we first consider the
representative class of two-qubit states
\begin{equation}\label{eq2-5}
 \tilde{\rho}= \frac{1}{4}\Bigl(I_4+\vec{r}\cdot\vec{\sigma}\otimes I_2
               +I_2\otimes\vec{s}\cdot\vec{\sigma}
               +\sum_{i=1}^3 v_{i}\sigma_i\otimes\sigma_i\Bigr),
\end{equation}
where $\{\vec{r},\vec{s},\vec{v}\}\in \mathbb{R}^3$ satisfy the
physical requirement $\tilde{\rho} \geqslant 0$. For $\vec{r}=
\vec{s}=0$, it reduces to the Bell-diagonal state
$\rho_{\mathrm{Bell}}$ which is characterized by the tetrahedron
$\mathcal{T}$ [see Fig. \ref{fig:1}(a)], and the region of separable
$\rho_{\mathrm{Bell}}$ is the octahedron $\mathcal{O}$
\cite{Horodecki}. For $\vec{r} \cdot \vec{s} \neq 0$, physical
$\tilde{\rho}$ shrinks to partial regions of $\mathcal{T}$. For this
case, while the separable region is still inside $\mathcal{O}$, the
entangled ones may not be limited to the four regions outside
$\mathcal{O}$.

\section{Hierarchy of NAQC and BN} \label{sec:3}
The hierarchy of entanglement, steering, and BN shows that while
entanglement clearly reveals the nonclassical nature of a state,
steering and BN exhibit even stronger deviations from classicality
\cite{new01,new02,hierarchy,Angelo1,Angelo2, Angelo3}. Here, we show
that NAQC may be viewed as a quantum correlation which is even
stronger than BN.

To begin with, we prove the convexity of NAQC,
\begin{equation}\label{eq3-1}
 C^{na}_\alpha\bigg({\sum_k q_k \rho_k}\bigg) \leqslant \sum_k q_k C^{na}_\alpha(\rho_k),
\end{equation}
that is, the NAQC is nonincreasing under mixing of states. By
combining Eqs. \eqref{eq2-2} and \eqref{eq2-3}, one can see that the
NAQC is convex provided $\bar{C}^{\sigma_j}_\alpha$ is convex. For
$\rho=\sum_k q_k \rho_k$, the conditional state of $B$ after Alice's
local measurements is
\begin{equation}\label{eq3-2}
 \rho_{B|\sigma_i^a}= \frac{\sum_k q_k \mathrm{tr}_A(\Pi_i^a\rho_k)}
                      {\sum_k q_k \mathrm{tr}(\Pi_i^a\rho_k)}
                    = \frac{\sum_k q_k p_k(a|\sigma_i)\rho_{B|\sigma_i^a}^k}{p(a|\sigma_i)},
\end{equation}
where $\rho_{B|\sigma_i^a}^k=\mathrm{tr}_A (\Pi_i^a \rho_k)/
p_k(a|\sigma_i)$, $p_k(a|\sigma_i)=\mathrm{tr}(\Pi_i^a \rho_k)$, and
we have denoted by $p(a|\sigma_i)=\sum_k q_k p_k(a|\sigma_i)$. Then
\begin{equation}\label{eq3-3}
  \begin{aligned}
   \bar{C}^{\sigma_j}_\alpha \big(\{p(a|\sigma_i), \rho_{B|\sigma_i^a}\}\big)
     &= \sum_a p(a|\sigma_i) C^{\sigma_j}_\alpha(\rho_{B|\sigma_i^a}) \\
     &\leqslant \sum_{k,a} p(a|\sigma_i) \frac{q_k p_k(a|\sigma_i)}{p(a|\sigma_i)}C^{\sigma_j}_\alpha\bigl(\rho_{B|\sigma_i^a}^k\bigr) \\
     &= \sum_{k,a} q_k p_k(a|\sigma_i) C^{\sigma_j}_\alpha\bigl(\rho_{B|\sigma_i^a}^k \bigr) \\
     &=\sum_k q_k \bar{C}^{\sigma_j}_\alpha\big(\{p_k(a|\sigma_i),\rho_{B|\sigma_i^a}^k\}\big),
  \end{aligned}
 \end{equation}
where the first inequality is due to convexity of the coherence
measure. This completes the proof of Eq. \eqref{eq3-1}.

\begin{figure}
\centering
\resizebox{0.39 \textwidth}{!}{%
\includegraphics{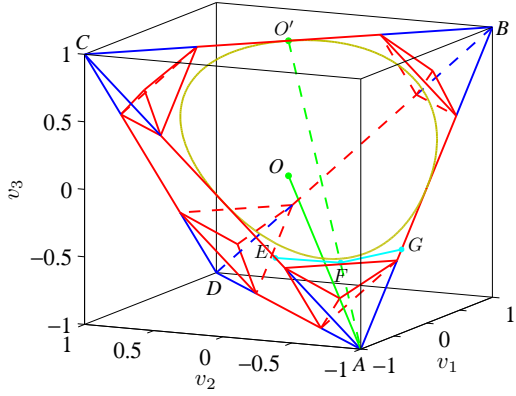}}
\caption{Geometric representation of the polyhedron $\mathcal {P}$
(the red lines) inside $\mathcal {T}$. Here, $O$ and $O'$ are the
coordinate origin and the midpoint of $BC$, respectively, while the
khaki curve is the curve of constant BN $M(\tilde{\rho})=1$ at the
facet $ABC$.} \label{fig:2}
\end{figure}

Next, we give the level surface $\mathcal{S}$ of constant BN
$M(\tilde{\rho})= 1$. It can be divided into four parts,
corresponding to the four vertices of $\mathcal{T}$. For convenience
of later presentation, we denote by $\mathcal{S}_A$ the part near
vertex $A$ (see Fig. \ref{fig:2}). It is described by
\begin{equation}\label{eq3-4}
 \begin{aligned}
 & v_i=\sin \theta,~ v_j=\cos \theta, \\
 & v_k\in[\max\{\sin\theta,\cos\theta\},1+\sin \theta+\cos \theta],
   ~ \theta\in[\pi, 1.5\pi],
\end{aligned}
\end{equation}
where $(i,j,k)=(1,2,3)$, $(2,3,1)$, and $(3,1,2)$. The equations for
the other three parts of $\mathcal {S}$ can be obtained directly by
their symmetry about the coordinate origin $O$. The corresponding
results are showed in Fig. \ref{fig:1}(b).

In the following, we denote by $\mathcal{N}$ the set of NAQC states
and $\mathcal{B}$ the set of Bell nonlocal states. We will prove the
inclusion relation $\mathcal {N}\subset\mathcal{B}$ for any
$\tilde{\rho}$, meaning that the existence of NAQC implies the
existence of BN.

\subsection{$l_1$ norm of of NAQC} \label{sec:3a}
First, we consider the class of Bell-diagonal states. Without loss
of generality, we assume $|v_1|\geqslant |v_2| \geqslant |v_3|$,
then
\begin{equation}\label{eq3a-1}
 C_{l_1}^{na}(\rho_{\mathrm{Bell}})=\sum_i|v_i|,~
 M(\rho_{\mathrm{Bell}})=v_1^2+v_2^2,
\end{equation}
from which one can obtain $|v_1|> \sqrt{6} /3$ and $|v_2|>(\sqrt{6}
-1)/2$ when $C_{l_1}^{na}(\rho_{\mathrm{Bell}})> \sqrt{6}$. This
further gives rise to $M(\rho_{\mathrm{Bell}})>1$. That is, any
$\rho_{\mathrm{Bell}}$ that can achieve a NAQC is Bell nonlocal. But
the converse is not true, e.g., if $v_{1,2,3} \in [-\sqrt{6}/3,
-1/\sqrt{2})$, we have $M(\rho_{\mathrm{Bell}})>1$ and $C_{l_1}^{na}
(\rho_{\mathrm{Bell}}) \leqslant \sqrt{6}$. With all this, we
arrived at the inclusion relation $\mathcal{N} \subset \mathcal{B}$.
The level surfaces of $C_{l_1}^{na} (\rho_{\mathrm{Bell}})=
\sqrt{6}$ can be found in Fig. \ref{fig:1}(c).

Second, we consider $\tilde{\rho}$ sitting at the edges of
$\mathcal{T}$ with general $\vec{r}$ and $\vec{s}$. We take the edge
$AB$ as an example (see Fig. \ref{fig:2}), the cases for the other
edges are similar. Along this edge, we have $v_1=v_3$ and $v_2=-1$,
then one can determine analytically the constraints imposed by
$\tilde{\rho}\geqslant 0$ on the involved parameters as $r_{1,3}=
s_{1,3}=0$, $r_2=-s_2$, and $s_2^2\leqslant 1-v_1^2$ (see Appendix
\ref{sec:A}). Thus we have
\begin{equation} \label{eq3a-2}
 C_{l_1}^{na}(\tilde{\rho}) = 1+|v_1|+\sqrt{v_1^2+s_2^2}.
\end{equation}
It is always not larger than $\sqrt{6}$ in the region of $|v_1|
\leqslant \sqrt{6}-2$. On the other hand, the states located at the
edge $AB$ other than its midpoint are Bell nonlocal. Hence, the
inclusion relation $\mathcal{N} \subset \mathcal{B}$ holds for all
$\tilde{\rho}$ located at the edges of $\mathcal{T}$.

Next, we consider $\tilde{\rho}$ associated with $v_{1,2,3}= v_0 =
-1/\sqrt{2}$. As $C_{l_1}^{na}$ is an increasing function of $|s_i|$
($i=1,2,3$), one only needs to determine the maximal $|s_i|$ for
which $\tilde{\rho}\geqslant 0$. Without loss of generality, we
assume $s_3= w_0 s_1$ and $s_2= w_1 s_1$, then a detailed analysis
shows that the resulting maximum NAQC states belong to the set of
$\tilde{\rho}$ with $r_3= w_0 r_1$ and $r_2= w_1 r_1$. Under this
condition, one can obtain analytically the eigenvalues $\epsilon_k$
of $\tilde{\rho}$. Then from $\epsilon_k \geqslant 0$ ($\forall k$)
one can obtain
\begin{equation}\label{eq3a-3}
 \begin{aligned}
  & |s_1+r_1|\leqslant c_1= \frac{1+v_0}{\sqrt{1+w_0^2+w_1^2}}, \\
  & |s_1-r_1|\leqslant c_2= \sqrt{\frac{1-2v_0-3v_0^2}{1+w_0^2+w_1^2}}.
 \end{aligned}
\end{equation}

For state $\tilde{\rho}$ with fixed $v_0$, $w_0$, and $w_1$,
$C_{l_1}^{na}$ takes its maximum when the above inequalities become
equalities. That is, when
\begin{equation}\label{eq3a-4}
 s_1=\pm\frac{1}{2}(c_1+c_2), ~
 r_1=\pm\frac{1}{2}(c_1-c_2),
\end{equation}
then by further maximizing the resulting $C_{l_1}^{na}$ over $w_0$
and $w_1$, we obtain $C_{l_1,\max}^{na} \simeq 2.4405$ at the
critical points $w_{0,1}=\pm 1$ (we have also checked the validity
of this result with $10^7$ randomly generated $\tilde{\rho}$ for
which $v_{1,2,3}= -1/\sqrt{2}$, and no violation was observed). As
this maximum is smaller than $\sqrt{6}$, any $\tilde{\rho}$ with
$v_{0}= -1/\sqrt{2}$ cannot achieve a NAQC.

To proceed, we introduce a polyhedron $\mathcal {P}$ with the set of
its vertices near the vertex $A$ being given by $(v_0,v_0,v_0)$,
$(-1,\gamma,\gamma)$, $(\gamma,-1,\gamma)$, $(\gamma,\gamma,-1)$,
and its other vertices can be obtained by using their symmetry with
respect to the point $O$ (see Fig. \ref{fig:2}). One can show that
when $|\gamma|< \sqrt{2}-1$, the surface $\mathcal {S}_A$ is always
inside $\mathcal{P}$ (see Appendix \ref{sec:B}). Finally, as
$C_{l_1,\max}^{na}\simeq 2.4405$ at the point $(v_0,v_0,v_0)$, we
choose $\gamma= 2-\sqrt{6}$ for which $C_{l_1}^{na}$ is also smaller
than $\sqrt{6}$ at the other three points of $\mathcal{P}$ near
vertex $A$ [see Eq.\eqref{eq3a-2}], then as any physical state with
$\vec{v}$ inside $\mathcal {P}$ can be written as a convex
combination of states with $\vec{v}$ at the vertices of $\mathcal
{P}$, we complete the proof of the inclusion relation $\mathcal{N}
\subset \mathcal{B}$ for general $\tilde{\rho}$ by using the
convexity of NAQC.

\begin{figure}
\centering
\resizebox{0.41 \textwidth}{!}{%
\includegraphics{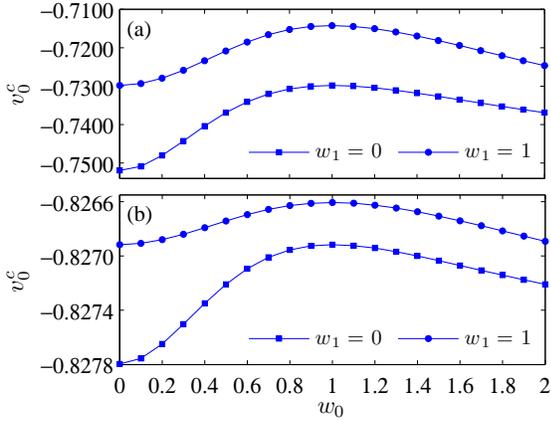}}
\caption{Critical $v_0^c$ for which (a) $C_{l_1}^{na}=\sqrt{6}$ and
(b) $C_{re}^{na}=C_{re}^m$ versus $w_0$ with $w_1=0$ and 1. The
other parameters are $\vec{s}=(s_1,w_0 s_1, w_1 s_1)$ and
$\vec{r}=(r_1,w_0 r_1, w_1 r_1)$, with $s_1$ and $r_1$ being given
in Eq. \eqref{eq3a-4}, and $v_0^c$ is plotted in the region of
$v_0<0$.} \label{fig:3}
\end{figure}

In fact, for $\tilde{\rho}$ at the line $AO$ with fixed $w_0$ and
$w_1$, one can obtain the critical $v_0^c$ at which $C_{l_1}^{na}=
\sqrt{6}$. As $C_{l_1}^{na}$ and $v_0^c$ considered here are
invariant under the substitution $w_0\leftrightarrow w_1$, we showed
in Fig. \ref{fig:3}(a) an exemplified plot of the $w_0$ dependence
of $v_0^c$ with fixed $w_1=0$ and 1. It first increases to a peak
value at $w_0=1$, then decreases gradually with the increase of
$|\omega_0|$. By optimizing over $w_0$ and $w_1$, one can further
obtain the region of $v_0^c \in (-0.7519, -0.7142)$, where the lower
and upper bounds correspond to $w_{0,1}=0$ and $w_{0,1}=\pm 1$,
respectively. Clearly, the point $(v_0^c,v_0^c,v_0^c)$ is always
outside the surface $\mathcal{S}$.

\subsection{Relative entropy of NAQC} \label{sec:3b}
In this subsection, we consider  NAQC measured by the relative
entropy. First, for Bell-diagonal states, the corresponding NAQC can
be obtained as \cite{naqc}
\begin{equation}\label{eq3b-1}
 C_{re}^{na}(\rho_{\mathrm{Bell}})= 3-\sum_i H\Biggl(\frac{1+v_i}{2}\Biggr).
\end{equation}
Then by imposing $C_{re}^{na}(\rho_{\mathrm{Bell}})>C_{re}^m$ with
the assumption $|v_1|\geqslant |v_2| \geqslant |v_3|$, one can
obtain
\begin{equation}\label{eq3b-2}
  H\Biggl(\frac{1+v_1}{2}\Biggr) < \frac{3-C_{re}^m}{3},~
  H\Biggl(\frac{1+v_2}{2}\Biggr) < \frac{3-C_{re}^m}{2},
\end{equation}
which yields $M(\rho_{\mathrm{Bell}}) > 1$. Moreover, we have
$M(\rho_{\mathrm{Bell}})>1$ and $C_{re}^{na}(\rho_{\mathrm{Bell}}) <
C_{re}^m$ for $v_{1,2,3} \in (-0.9140, -1/\sqrt{2})$. So
$\mathcal{N} \subset \mathcal{B}$ holds for $\rho_\mathrm{Bell}$.
The corresponding level surfaces were showed in Fig. \ref{fig:1}(d).
Clearly, the region of NAQC states shrinks compared with that
captured by the $l_1$ norm.

For $\tilde{\rho}$ sitting at the edges of $\mathcal {T}$ with
general $\vec{r}$ and $\vec{s}$, we take the edge $AB$ as an
example. Based on the results of Sec. \ref{sec:3a}, one can obtain
\begin{equation}\label{eq3b-3}
 C_{re}^{na}= 2+H\Biggl(\frac{1+s_2}{2}\Biggr)-
              2H\left(\frac{1+\sqrt{v_1^2+s_2^2}}{2}\right),
\end{equation}
then it is direct to show that $C_{re}^{na}$ is always smaller than
$C_{re}^m$ for $|v_1|<-b_0\simeq 0.3813$. So the inclusion relation
$\mathcal{N}\subset \mathcal{B}$ holds for any $\tilde{\rho}$ at the
edges of $\mathcal {T}$.

Based on the above preliminaries, we now consider $\tilde{\rho}$ at
the surface $\mathcal {S}_A$ (the cases for the other parts of
$\mathcal {S}$ are similar). We will show that for these
$\tilde{\rho}$ the inequality $C_{re}^{na} < C_{re}^m$ holds. Then
by further employing the convexity of NAQC and the fact that
$\{\tilde{\rho}\}$ is a convex set, one can complete the proof of
$\mathcal{N} \subset \mathcal{B}$. In fact, due to the structure of
$\mathcal {S}_A$ [see Eq. \eqref{eq3-4}], it suffices to prove that
we always have $C_{re}^{na} < C_{re}^m$ at the boundary of
$\mathcal{S}_A$.

First, we introduce the polygon line $EFG$ over $(-1,b_0,b_0)$,
$(a_0, a_0, 1+2a_0)$, and $(b_0,-1, b_0)$. One can prove that there
is no intersection of this line and the boundary of $\mathcal{S}_A$
at the facet $ABC$ when $a_0\lesssim -0.7082$ (Appendix
\ref{sec:B}). Moreover, along the line $AO'$, $\tilde{\rho}\geqslant
0$ yields $r_{1,2}= -s_{1,2}$ and $r_3=s_3$ (Appendix \ref{sec:A}),
then one can obtain that at the point $F$ with $a_0= -0.7082$,
$C_{re}^{na}$ maximized over $\vec{r}$ and $\vec{s}$ is of about
1.4956. As $C_{re}^{na}$ is also smaller than $C_{re}^m$ at the
points $E$ and $G$ [see Eq. \eqref{eq3b-3}], we have $C_{re}^{na}<
C_{re}^m$ for any $\tilde{\rho}$ at this boundary.

Second, if we make the substitutions $v_0=-0.7082$ and $\gamma=b_0$
to the vertices of $\mathcal {P}$, then one can show that the
boundary of $\mathcal{S}_A$ inside $\mathcal {T}$ is also inside
$\mathcal{P}$ (see Appendix \ref{sec:B}). For $\tilde{\rho}$ at the
point $(v_0,v_0,v_0)$, our numerical results showed that with fixed
$v_0$, $w_0$, and $w_1$, $C_{re}^{na}$ also takes its maximum when
$s_1$ and $r_1$ are given by Eq. \eqref{eq3a-4}. Then by further
maximizing it over $w_0$ and $w_1$, we obtain $C_{re,\max}^{na}
\simeq 2.0041$ at $w_{0,1}=\pm 1$. As $C_{re}^{na}$ is also smaller
than $C_{re}^m$ for $\tilde{\rho}$ at the vertices of $\mathcal {P}$
with $\gamma=b_0$ [see Eq. \eqref{eq3b-3}], we have $C_{re}^{na}<
C_{re}^m$ for any $\tilde{\rho}$ at this boundary.

Similar to the $l_1$ norm of NAQC, one can obtain $v_0^c$ at which
$C_{re}^{na}= C_{re}^m$ with fixed $w_0$ and $w_1$. It is $v_0^c\in
(-0.8278, -0.8266)$, where the lower and upper bounds are obtained
with $w_{0,1}=0$ and $w_{0,1}=\pm 1$, respectively. As is showed in
Fig. \ref{fig:3}, $v_0^c$ for the two NAQCs exhibits qualitatively
the same $w_0$ dependence.

Before ending this section, we would like to mention here that
although for the set of Bell-diagonal states, one detects a wider
region of NAQC states by using the $l_1$ norm as a measure of
coherence than that by using the relative entropy (see Fig.
\ref{fig:1}), this is not always the case. A typical example is that
for $\tilde{\rho}$ at the edge $AB$ of $\mathcal {T}$ with
$|v_{1,3}|\in (0.3813, \sqrt{6}-2)$, one may have $C_{l_1}^{na}<
\sqrt{6}$ and $C_{re}^{na}>C_{re}^m$.

\section{An explicit application of NAQC} \label{sec:4}
As it is a proven fact that all Bell nonlocal states are useful for
quantum teleportation \cite{telep}, the hierarchy we obtained
implies that any NAQC state $\tilde{\rho}$ can serve as a quantum
channel for quantum teleportation. That is, it always gives rise to
the average fidelity $F_{av}>2/3$. In fact, $F_{av}$ achievable with
the channel state $\tilde{\rho}$ is given by \cite{telep}
\begin{equation}\label{eq4-1}
 F_{av}(\tilde{\rho})=\frac{1}{2}+ \frac{1}{6} \sum_i |v_i|.
\end{equation}
Using this equation and the results of Sec. \ref{sec:3}, one can
obtain that for any NAQC state $\tilde{\rho}$ captured by
$C_{l_1}^{na}(\tilde{\rho})$, we always have $F_{av}> \sqrt{6}/3$,
while for any NAQC state $\tilde{\rho}$ captured by $C_{re}^{na}
(\tilde{\rho})$, we always have $F_{av}\gtrsim 0.7938$. Both the two
critical values are larger than $2/3$, so any NAQC state
$\tilde{\rho}$ can serve as a quantum channel for nonclassical
teleportation.

If we focus only on the class of NAQC Bell-diagonal states, the
average fidelity $F_{av}$ can be further improved. More
specifically, Eqs. \eqref{eq3a-1} and \eqref{eq3b-1} imply that
$F_{av}>(3+\sqrt{6})/6$ for any NAQC state $\rho_\mathrm{Bell}$
captured by $C_{l_1}^{na}(\rho_\mathrm{Bell})$, and $F_{av} \gtrsim
0.9501$ for any NAQC state $\rho_\mathrm{Bell}$ captured by
$C_{re}^{na} (\rho_\mathrm{Bell})$.

\section{Summary and discussion} \label{sec:5}
In summary, we have explored the interrelations of NAQC achievable
in a two-qubit state under local measurements and BN detected by
violation of the Bell-CHSH inequality. There are two different
scenarios of NAQC being considered: one is characterized by the
$l_1$ norm of coherence, and another one is characterized by the
relative entropy of coherence. For both scenarios, we showed
geometrically that the inclusion relation $\mathcal{N} \subset
\mathcal{B}$ holds for the class of states $\tilde{\rho}$ that have
diagonal correlation matrix $T$. This extends the known hierarchy in
quantum correlation, viz., BN, steerability, entanglement, and
quantum discord to include NAQC.

One may also concern whether the obtained hierarchy holds for $\rho$
with nondiagonal $T$. As such $\rho$ is locally unitary equivalent
to $\tilde{\rho}$, that is, $\rho=U_{AB} \tilde{\rho} U_{AB}^\dag$
with $U_{AB}=U_A\otimes U_B$, the proof can be completed by showing
that for any $\tilde{\rho}$ with $M(\tilde{\rho})\leqslant 1$, we
have $C_{\alpha}^{na} (U_{AB} \tilde{\rho} U_{AB}^\dag) \leqslant
C_{\alpha}^m$ for all unitaries $U_{AB}$. But due to the so many
number of state parameters involved, it is difficult to give such a
strict proof. For special cases, a strict proof may be available,
e.g., for the locally unitary equivalent class of $\tilde{\rho}$
with $|\vec{v}|^2+2|\vec{s}|^2 \leqslant 2$, we are sure that
$C_{l_1}^{na}\leqslant\sqrt{6}$, while for the locally unitary
equivalent class of $\rho_\mathrm{Bell}$, we are sure that
$C_{re}^{na} \leqslant C_{re}^m$ (see Appendix \ref{sec:C}).
Moreover, for $\tilde{\rho}$ with reduced number of parameters, we
performed numerical calculations with $10^7$ equally distributed
local unitaries generated according to the Haar measure \cite{Haar1,
Haar2}, and found that $C_{\alpha}^{na}$ is always smaller than
$C_\alpha^m$ (see Appendix \ref{sec:C}). These results presented
strong evidence that the hierarchy may hold for any two-qubit state,
though a strict proof is still needed.

Moreover, one may argue that NAQC can be recognized as a quantum
correlation. It is stronger than BN in the sense that the NAQC
states form a subset of the Bell nonlocal states. But it is
asymmetric, that is, in general $C^{na}_\alpha$ defined with the
local measurements on $A$ does not equal that defined with the local
measurements on $B$. This property is the same to steerability and
quantum discord. The NAQC is also not locally unitary invariant. Its
value may be changed by performing local unitary transformation to
the mutually unbiased bases. To avoid this perplexity, one can
define
\begin{equation}\label{eq5-1}
 \tilde{C}_\alpha^{na}(\rho)= \frac{1}{2} \max_{\{U_A\otimes U_B\}} \sum_{i,j,a\atop i\neq j}
                     p(a|\sigma_{i,U_A})C^{\sigma_{j,U_B}}_\alpha (\rho_{B|\sigma_{i,U_A}^a}),
\end{equation}
with $\sigma_{i,U_A} = U_A\sigma_i U_A^\dag$, and likewise for
$\sigma_{j,U_B}$. As BN is locally unitary invariant, we have
$\tilde{\mathcal{N}} \subset \mathcal {B}$ provided $\mathcal{N}
\subset \mathcal {B}$, where $\tilde{\mathcal {N}}$ is the set of
NAQC states captured by $\tilde{C}_\alpha^{na} (\rho)> C_\alpha^m$ .

Finally, in light of those measures of steerability based on the
maximal violation of various steering inequalities and the similar
measure of Bell nonlocality \cite{Angelo1,Angelo2}, it is natural to
quantify the degree of NAQC in a bipartite state $\rho$ by
\begin{equation}\label{eq5-2}
 \tilde{Q}_{\alpha}(\rho)= \max\left\{0,\frac{\tilde{C}_\alpha^{na}(\rho)-C_\alpha^m}
                           {\tilde{C}_{\alpha,\max}^{na}- C_\alpha^m}\right\},
\end{equation}
where $\tilde{C}_{\alpha,\max}^{na}= \max_\rho \tilde{C}_\alpha^{na}
(\rho)$, and  the factor $\tilde{C}_{\alpha,\max}^{na}- C_\alpha^m$
was introduced for normalizing $\tilde{Q}_{\alpha}(\rho)$. For
two-qubit states, we have $\tilde{C}_{\alpha,\max}^{na}=3$
($\alpha=l_1$ or $re$), which are obtained for the Bell states
$|\Phi^\pm\rangle=(|00\rangle\pm|11\rangle)/\sqrt{2}$ and
$|\Psi^\pm\rangle=(|01\rangle\pm|10\rangle)/\sqrt{2}$. Moreover, we
have used the fact that $C_\alpha^m$ cannot be increased by any
unitary transformation in the above definition.

\begin{figure}
\centering
\resizebox{0.41 \textwidth}{!}{%
\includegraphics{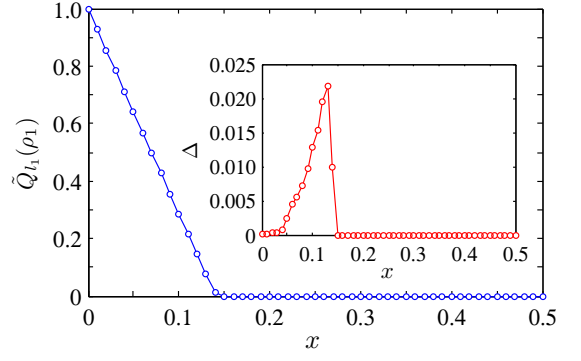}}
\caption{The $x$ dependence of $\tilde{Q}_{l_1}(\rho_1)$. Here, we
choose $x\in [0,0.5]$ as $\tilde{Q}_{l_1}(\rho_1)$ is symmetric with
respect to $x=0.5$ for $\rho_1$. The inset shows the $x$ dependence
of $\Delta=\tilde{Q}_{l_1}(\rho_1)-Q_{l_1}(\rho_1)$.} \label{fig:4}
\end{figure}

Of course, one may propose to define the NAQC-based correlation
measure [denoted $Q_{l_1}(\rho)$] by replacing
$\tilde{C}_\alpha^{na}(\rho)$ in Eq. \eqref{eq5-2} with
$C_\alpha^{na}(\rho)$. But if so, $Q_{l_1}(\rho)$ will not be
locally unitary invariant, thus makes it violates the widely
accepted property of a quantum correlation measure (e.g.,
 Bell nonlocality, steerability, entanglement, and quantum discord)
which should be locally unitary invariant.

As an example, we calculated numerically the NAQC-based correlation
measure of the following state
\begin{equation}\label{eq5-3}
 \rho_1=x|\Phi^+\rangle\langle\Phi^+|+(1-x)|\Psi^-\rangle\langle\Psi^-|,~~ x\in[0,1],
\end{equation}
for which $\tilde{Q}_{l_1}(\rho_1)$ is symmetric with respect to
$x=0.5$. As was showed in Fig. \ref{fig:4}, $\tilde{Q}_{l_1}
(\rho_1)> Q_{l_1}(\rho_1)$ in the region of $0\lesssim x\lesssim
0.141$. In particular, we have $\tilde{Q}_{l_1}(\rho_1) >0$ and
$Q_{l_1}(\rho_1)=0$ when $0.138 \lesssim x\lesssim 0.141$, that is,
$\tilde{Q}_{l_1}$ captures a wider region of NAQC states than
$Q_{l_1}$.

\section*{ACKNOWLEDGMENTS}
This work was supported by National Natural Science Foundation of
China (Grants No. 11675129, No. 91536108, and No. 11774406),
National Key R \& D Program of China (Grants No. 2016YFA0302104 and
No. 2016YFA0300600), the New Star Project of Science and Technology
of Shaanxi Province (Grant No. 2016KJXX-27), the Strategic Priority
Research Program of Chinese Academy of Sciences (Grant No.
XDB28000000), and the New Star Team of XUPT.

\begin{appendix}
\section{Constraints imposed on the parameters of $\tilde{\rho}$} \label{sec:A}
\setcounter{equation}{0}
\renewcommand{\theequation}{A\arabic{equation}}
\setcounter{figure}{0}
\renewcommand{\thefigure}{A\arabic{figure}}
At the edge $AB$ of $\mathcal {T}$, we have $v_1=v_3$ and $v_2=-1$.
Then the positive semidefiniteness of $\tilde{\rho}$ requires
\begin{equation}\label{eqa-1}
 \begin{aligned}
  & \tilde{\rho}_{11}\tilde{\rho}_{44}- |\tilde{\rho}_{14}|^2=-(r_3+s_3)^2\geqslant 0,\\
  & \tilde{\rho}_{22}\tilde{\rho}_{33}- |\tilde{\rho}_{23}|^2=-(r_3-s_3)^2\geqslant 0,
 \end{aligned}
\end{equation}
from which one can obtain $r_3=s_3=0$.

Moreover, all the {\textit i}th-order principal minors of
$\tilde{\rho}$ should be nonnegative. Under the constraint
$r_3=s_3=0$ obtained above, the second- and third-order leading
principal minors $D_{2,3}$ and the principal minor $\Delta_3$
(determinant of the matrix obtained by removing from $\tilde{\rho}$
its third row and third column) are
\begin{equation}\label{eqa-2}
 \begin{aligned}
  & D_2=1-v_1^2-s_1^2-s_2^2, \\
  & D_3=(v_1-1)[(r_1+s_1)^2+ (r_2+s_2)^2], \\
  & \Delta_3=- (v_1+1)[(r_1-s_1)^2+(r_2+s_2)^2],
 \end{aligned}
\end{equation}
which, together with Eq. \eqref{eqa-1}, yields the following
requirements
\begin{equation}\label{eqa-3}
  r_{1,3}=s_{1,3}=0,~ r_2=-s_2,~ s_2^2\leqslant 1-v_1^2.
\end{equation}

Similarly, one can obtain constraints imposed on the parameters of
$\tilde{\rho}$ at the other edges of $\mathcal {T}$. They are
\begin{equation}\label{eqa-4}
 \begin{aligned}
  &AC\mathrm{:}~ r_{2,3}=s_{2,3}=0,~ r_1=-s_1,~ s_1^2\leqslant 1-v_3^2, \\
  &AD\mathrm{:}~ r_{1,2}=s_{1,2}=0,~ r_3=-s_3,~ s_3^2\leqslant 1-v_2^2, \\
  &CD\mathrm{:}~ r_{1,3}=s_{1,3}=0,~ r_2=s_2,~  s_2^2\leqslant 1-v_1^2, \\
  &BD\mathrm{:}~ r_{2,3}=s_{2,3}=0,~ r_1=s_1,~  s_1^2\leqslant 1-v_3^2, \\
  &BC\mathrm{:}~ r_{1,2}=s_{1,2}=0,~ r_3=s_3,~  s_3^2\leqslant 1-v_2^2.
 \end{aligned}
\end{equation}

For $\tilde{\rho}$ associated with $\vec{v}$ at the line $AO'$, we
have $v_{1,2}=a_0$ and $v_3=1+2a_0$ ($-1\leqslant a_0\leqslant 0$),
then a similar derivation gives
\begin{equation}\label{eqa-5}
 \begin{aligned}
  & r_{1,2}=-s_{1,2},~ r_3=s_3\in[-1-a_0, 1+a_0], \\
  & |s_1|\leqslant \min\Big\{1+\frac{1}{2}a_0, \frac{1}{2}(1-a_0)\Big\},\\
  & s_1^2+s_2^2 \leqslant -4a_0(1+a_0).
 \end{aligned}
\end{equation}

\section{Intersection of two surfaces} \label{sec:B}
\setcounter{equation}{0}
\renewcommand{\theequation}{B\arabic{equation}}
Due to the symmetry, one only needs to consider the intersections of
the level surface $\mathcal {S}_A$ described by Eq. \eqref{eq3-4}
and the facet of $\mathcal {P}$ with the vertices $(v_0,v_0, v_0)$,
$(-1,\gamma, \gamma)$, $(\gamma,-1,\gamma)$. The plane equation for
this facet is
\begin{equation}\label{eqb-1}
 a v_1+a v_2+c v_3+1=0,
\end{equation}
where
\begin{equation}\label{eqb-2}
 a=\frac{v_0-\gamma}{v_0(1+\gamma)},~c=-\frac{1}{v_0}-2a.
\end{equation}

Without loss of generality, we fix $(i,j,k)=(1,2,3)$ in Eq.
\eqref{eq3-4}. Then by plugging $v_1=\sin\theta$ and
$v_2=\cos\theta$ into Eq. \eqref{eqb-1}, we obtain
\begin{equation}\label{eqb-3}
 v_3=\frac{(v_0-\gamma)(\sin\theta+\cos\theta)+v_0(1+\gamma)}
          {1+2v_0-\gamma},
\end{equation}
and for given $v_0$ and $\gamma$, one can check whether there are
intersections for the two surfaces by checking whether $v_3$
obtained in Eq. \eqref{eqb-3} belongs to the region
$[\max\{\sin\theta, \cos\theta\}, 1+\sin\theta+\cos\theta]$. If
there exists such $v_3$, then there are intersection of
$\mathcal{S}_A$ and $\mathcal{P}$. Otherwise, $\mathcal {S}_A$ is
totally inside or outside of $\mathcal{P}$.

One can also determine whether there are intersections of
$\mathcal{S}_A$ and $\mathcal {P}$ by plugging Eq. \eqref{eq3-4}
into Eq. \eqref{eqb-1}, and checking the resulting $\mathrm{sgn}( a
v_1+a v_2+c v_3+1)$. The surface $\mathcal{S}_A$ is inside
$\mathcal {P}$ if it is always nonnegative. In fact, here one only
needs to check the points at the boundary of $\mathcal {S}_A$.

Based on the above methods, it is direct to show that when
$v_0=-1/\sqrt{2}$ and $|\gamma|<\sqrt{2}-1$, the level surface
$\mathcal {S}_A$ is always inside $\mathcal{P}$. When $v_0\lesssim
-0.7082$ and $\gamma= b_0$, the boundary of $\mathcal {S}_A$ inside
the tetrahedron $\mathcal{T}$ is also inside the polyhedron
$\mathcal{P}$.

Similarly, by substituting $v_1=\sin\theta$, $v_2=\cos\theta$, and
$v_3=1+\sin\theta+\cos\theta$ into the equation of the straight line
$FG$ (see Fig. \ref{fig:2}), one can obtain
\begin{equation}\label{eqb-4}
(1+a_0)\sin\theta+(b_0-a_0)\cos\theta=a_0(1+b_0).
\end{equation}
For given $a_0$ and $b_0$, if there are solutions for Eq.
\eqref{eqb-4} in the region of $\theta\in [\pi,1.5\pi]$, there are
intersections of $FG$ and the boundary of $\mathcal{S}_A$ described
by $v_3=1+\sin\theta+ \cos\theta$. In this way, one can check that
when $b_0\simeq -0.3813$ and $a_0\lesssim -0.7082$, there are no
intersections of $FG$ and the boundary of $\mathcal{S}_A$.

\section{NAQC of general two-qubit states} \label{sec:C}
\setcounter{equation}{0}
\renewcommand{\theequation}{C\arabic{equation}}
Suppose $U_{AB}= U_A \otimes U_B$ gives the map $\vec{r} \mapsto
\vec{x}$, $\vec{s} \mapsto \vec{y}$, and $\vec{v} \mapsto
T=(t_{ij})$, then the transformed state of $\tilde{\rho}$ is given
by
\begin{equation}\label{eqc-1}
 \rho= \frac{1}{4}\Big(I_4+\vec{x}\cdot\vec{\sigma}\otimes I_2
       +I_2\otimes\vec{y}\cdot\vec{\sigma}
       +\sum_{i,j=1}^3 t_{ij}\sigma_i\otimes\sigma_j\Big),
\end{equation}
and we have the following equalities
\begin{equation}\label{eqc-2}
 |\vec{r}|=|\vec{x}|,~ |\vec{s}|=|\vec{y}|,~
 |\vec{v}|^2=\sum_{ij}t_{ij}^2.
\end{equation}

By further using the mean inequality and the analytical solution of
$C_{l_1}^{na}(\rho)$ given in Ref. \cite{naqc}, we obtain
\begin{equation}\label{eqc-3}
 \begin{aligned}
  C_{l_1}^{na}(U_{AB}\tilde{\rho} U_{AB}^\dag)
         & \leqslant \sqrt{\frac{3}{2}\Big(|\vec{v}|^2+ \sum_{i} t_{ii}^2 \Big)+6|\vec{s}|^2}, \\
         & \leqslant \sqrt{3|\vec{v}|^2+6|\vec{s}|^2},
 \end{aligned}
\end{equation}
hence for the class of $\tilde{\rho}$ with $|\vec{v}|^2+2|\vec{s}|^2
\leqslant 2$, we are sure that $C_{l_1}^{na}(U_{AB}\tilde{\rho}
U_{AB}^\dag) \leqslant \sqrt{6}$. This class of $\tilde{\rho}$
includes (but not limited to) all $\tilde{\rho}$ with
$|\vec{s}|^2\leqslant 1/4$ as we have $|\vec{v}|^2\leqslant 3/2$ for
$M(\tilde{\rho})\leqslant 1$.

For the relative entropy of NAQC, due to its complexity, we consider
only the case of $\rho_\mathrm{Bell}$, for which we have
\begin{equation}\label{eqc-4}
 \begin{aligned}
  C_{re}^{na}(U_{AB} \rho_\mathrm{Bell} U_{AB}^\dag)
      =& \frac{1}{2}\sum_{i\neq j}H \Biggl(\frac{1+t_{ij}}{2}\Biggr) \\
       & -\sum_i H \left(\frac{1+\sqrt{\sum_j t_{ij}^2}}{2} \right),
 \end{aligned}
\end{equation}
then by using $|\vec{v}|^2\leqslant 3/2$ when $M(\tilde{\rho})
\leqslant 1$, one can show that the maximum of the right-hand side
of Eq. \eqref{eqc-4} is of about 1.1974, which is achieved when
$T=\mathrm{diag}\{v_0,v_0,v_0\}$, with $v_0=- 1/\sqrt{2}$. Hence
$C_{re}^{na}(U_{AB} \rho_\mathrm{Bell} U_{AB}^\dag)< C_{re}^m$ for
this class of $\tilde{\rho}$.

For general $\tilde{\rho}$ inside the level surface $\mathcal {S}$,
it is hard even to give a numerical simulation as the derivation of
the constraints imposed on $\vec{r}$ and $\vec{s}$ is also a
difficult task. But if the number of the involved parameters can be
reduced, a numerical verification may also be possible. Several
examples where such a verification can be performed are as follows:

(1) For the class of $\tilde{\rho}$ at the vertex $(0,-1,0)$ of
$\mathcal {O}$ (the cases for the other vertices of $\mathcal {O}$
are similar), we have $r_{1,3}=s_{1,3}=0$ and $r_2=-s_2$, i.e.,
there is only one variable. We performed numerical calculation with
$10^7$ equally distributed local unitaries generated according to
the Haar measure \cite{Haar1,Haar2}, and found that the maximal
$C_{l_1}^{na}$ and $C_{re}^{na}$ achievable by optimizing over
$U_A\otimes U_B$ increase with the increase of $|s_2|$. When
$|s_2|=1$, their maximal values are $\sqrt{6}$ and $C_{re}^m$,
respectively. The corresponding optimal $U_{AB}\tilde{\rho}
U_{AB}^\dag$ is of the form of Eq. \eqref{eqc-1}, with
\begin{equation} \label{eqc-5}
 \vec{x}=-\vec{y}=\biggl(\pm \frac{1}{\sqrt{3}},\pm \frac{1}{\sqrt{3}},
                         \pm \frac{1}{\sqrt{3}}\biggr),~~
 t_{ij}=-\frac{1}{3}~(\forall i,j).
\end{equation}

(2) For the class of $\tilde{\rho}$ associated with $v_{1,2,3}=
-1/\sqrt{2}$ (the cases for $v_i=v_j=-v_k=1/\sqrt{2}$ are similar),
the parameter regions can be reduced via $r_3^2+s_3^2 \leqslant
1-v_1^2$ and $|r_{1,3}\pm s_{1,3}|\leqslant 1\pm v_1$. The numerical
results show that $C_{\alpha}^{na}(U_{AB} \tilde{\rho}U_{AB}^\dag)$
is still smaller than $C_{\alpha}^m$ ($\alpha=l_1$ or $re$).
Specifically, when $w_{0,1}=\pm 1$, $s_1$ and $r_1$ take the values
of Eq. \eqref{eq3a-4}, the NAQC of $\tilde{\rho}$ cannot be enhanced
by $U_{AB}$, i.e., $C_{l_1}^{na}(\tilde{\rho}) \simeq 2.4405$ and
$C_{re}^{na} (\tilde{\rho})\simeq 2.0026$ are already the maximum
values.

(3) For the class of $\tilde{\rho}$ with $v_{1,2}=-1/\sqrt{2}$ and
$v_3=1-\sqrt{2}$ [an intersection of $AO'$ and the curve of
$M(\tilde{\rho})=1$], one can obtain $|\vec{s}|^2\leqslant
\sqrt{2}-1$ by using Eq. \eqref{eqa-5}. Hence $|\vec{v}|^2+
2|\vec{s}|^2 < 2$, and $C_{l_1}^{na}(U_{AB} \tilde{\rho}
U_{AB}^\dag)$ cannot exceed $\sqrt{6}$ due to Eq. \eqref{eqc-3}. For
NAQC characterized by the relative entropy, we performed numerical
calculation with $10^3$ equally distributed $\tilde{\rho}$ of this
class, while every $\tilde{\rho}$ is further optimized over $10^7$
equally distributed local unitaries. From these calculation we still
have not found the case for which $C_{re}^{na} (U_{AB} \tilde{\rho}
U_{AB}^\dag)>C_{re}^m$.

\end{appendix}

\newcommand{\PRL}{Phys. Rev. Lett. }
\newcommand{\RMP}{Rev. Mod. Phys. }
\newcommand{\PRA}{Phys. Rev. A }
\newcommand{\PRB}{Phys. Rev. B }
\newcommand{\PRE}{Phys. Rev. E }
\newcommand{\PRX}{Phys. Rev. X }
\newcommand{\NJP}{New J. Phys. }
\newcommand{\JPA}{J. Phys. A }
\newcommand{\JPB}{J. Phys. B }
\newcommand{\PLA}{Phys. Lett. A }
\newcommand{\NP}{Nat. Phys. }
\newcommand{\NC}{Nat. Commun. }
\newcommand{\SR}{Sci. Rep. }
\newcommand{\EPJD}{Eur. Phys. J. D }
\newcommand{\QIP}{Quantum Inf. Process. }
\newcommand{\QIC}{Quantum Inf. Comput. }
\newcommand{\AoP}{Ann. Phys. }
\newcommand{\PR}{Phys. Rep. }
%

%

\end{document}